\documentclass[11pt,a4paper]{article}

\usepackage{amsmath,amsthm,mathtools,url,amssymb,upgreek,textgreek,slashed,hyperref}
\usepackage{soul,color}
\usepackage{cite}
\usepackage[mathscr]{eucal}
  \usepackage[pdftex]{graphicx}
  \usepackage{epstopdf}
\allowdisplaybreaks
\begin{document}
\numberwithin{equation}{section}
\begin{titlepage}

\vskip 1.5in
\begin{center}
{\bf{Generalized Half-Dyon in SU(2) Yang-Mills-Higgs Theory}}\\ \vskip.5cm

\vskip 0.5cm  {Guo-Quan Wong, Khai-Ming Wong, and Dan Zhu} \vskip 0.15in {\small{ \textit{School of Physics}\vskip .1cm
{\textit{Universiti Sains Malaysia, 11800 USM Penang, Malaysia}}}
}
\end{center}
\vskip 0.5in
\baselineskip 16pt
\begin{abstract}   
We report on generalized half-dyon solutions in SU(2) Yang-Mills-Higgs theory, namely Type I and Type II solutions. These solutions are constructed by considering $\phi$-winding number $1\leq n \leq 4$, electric charge parameter $0 \leq\eta <\eta_{max}$ and Higgs self-coupling constant $0 \leq \beta \leq 1$. They represent system of magnetic charge $+2n\pi/g$ ($-2n\pi/g$) and electric charge $Q$ that lies along the negative (positive) $z$-axis. Fundamental properties such as total energy, electric charge, dipole moment are explored and discussed.
\end{abstract}
\date{May, 2023}
\end{titlepage}

\section{Introduction}\label{intro}
In Maxwell theory, a monopole is accompanied by a Dirac string, and the Dirac monopole \cite{kn:1} possessing infinite energy obeys the quantization condition $2Me/\hbar c=N$, where $M$ is the magnetic charge, $e$ is the U(1) gauge field coupling constant, and $N$ is an integer. However, in the presence of a Higgs field in the adjoint representation, the 't Hooft-Polyakov monopole \cite{kn:2,kn:3} arises in the non-Abelian SU(2) Yang-Mills field and the Dirac string disappears, resulting in a finite energy monopole with magnetic charge one that is radially symmetrical. The quantization condition in the SU(2) Yang-Mills-Higgs field theory in $3+1$ dimensions becomes $Mg/\hbar c=N$, where $g$ is the SU(2) gauge field coupling constant.

In $3+1$ dimensions, the SU(2) Yang-Mills-Higgs (YMH) field theory with the Higgs field in the adjoint representation possesses magnetic monopole configurations, as shown in Refs.~\citen{kn:2,kn:3,kn:4,kn:5,kn:6,kn:7,kn:8,kn:9,kn:10,kn:11,kn:12}. Among these, the 't Hooft-Polyakov monopole solution is invariant under a U(1) subgroup of the local SU(2) gauge group, has non-zero Higgs mass and self-interaction, and has been found to be spherically symmetric and possess finite energy \cite{kn:2,kn:3}. However, numerical solutions for monopole configurations with magnetic charges greater than unity cannot possess spherical symmetry, as demonstrated in Refs.~\citen{kn:8,kn:11,kn:13}. Exact monopole solutions exist only in the Bogomol'nyi-Prasad-Sommerfield (BPS) limit \cite{kn:5,kn:6,kn:7,kn:9,kn:12}, and only numerical solutions have been found when the Higgs field potential is non-vanishing. Other exact solutions include the A-M-A and vortex ring solutions, as well as various mirror symmetric monopole configurations \cite{kn:14,kn:15}, and numerical BPS monopole solutions with no rotational symmetry have also been discussed \cite{kn:16,kn:17}. Further numerical finite energy monopole solutions include the monopole-antimonopole pair (MAP), monopole-antimonopole chain (MAC), and vortex ring solutions \cite{kn:18,kn:19,kn:20,kn:21}.

While most of the literature on monopole solutions focuses on those with integer topological magnetic charge, there have been some discussions on half-monopoles in recent years. Harikumar et al.~\cite{kn:22} demonstrated the existence of generic smooth Yang-Mills (YM) potentials of one-half monopoles, but no exact or numerical solutions were provided. Exact axially symmetric one-half monopole solutions and mirror symmetric one-half monopole solutions with Dirac-like strings were discussed in Ref.~\citen{kn:16,kn:17}, but these solutions possess infinite total energy. Teh et al.~\cite{kn:23} recently found a finite energy one-half monopole solution. The 't Hooft magnetic fields of these solutions at spatial infinity correspond to the magnetic field of a positive one-half magnetic monopole located at the origin ($r=0$) and a semi-infinite Dirac string located on one half of the $z$-axis carrying magnetic flux, $2\pi/g$, going into the center of the sphere at infinity.

Dyon refers to a particle possessing both magnetic and electric charges. A dyon with a fixed magnetic charge can possess varying electric charges \cite{kn:24} at the classical level. The dyon solutions of Julia and Zee \cite{kn:25,kn:26} are time-independent solutions with non-vanishing kinetic energy. However, they are non-self-dual even in the BPS limit when the electric charge is non-vanishing. Outside the BPS limit, when $\lambda$ is non-vanishing, the Julia-Zee solutions are non-self-dual. The exact dyon solutions found by Prasad and Sommerfield \cite{kn:5} are, in fact, Julia and Zee dyon solutions in the BPS limit. Teh et al.~\cite{kn:27} also found a one-half dyon with finite energy for the case of $\phi$-winding number $n=1$. They reported on its properties, and the one-half dyon solution does not satisfy the first order Bogomol'nyi equations.

This paper presents a construction of axially symmetric half-dyon solutions in YMH theory by extending the previous work conducted by Teh et al.~\cite{kn:27}. We consider the case with $\phi$-winding number, $1 \leq n \leq 4$ with electric charge parameter, 0 $\leq \eta < \eta_{max} $ and the Higgs self-coupling constant, $ 0 \leq \beta \leq 1$. Specifically, our attention is directed towards $\beta = 0.7782$, since it corresponds to the physical value of $\beta$ in Weinberg-Salam model.

Additionally, we construct generalized class of half-dyon solutions, denoted as Type I and Type II. The Type I solution refers to the half-dyon configuration characterized by positive magnetic charge extending along the negative $z$-axis. These Type I solution has been partially discussed in Ref.~\citen{kn:27} and they are further studied in this paper. Conversely, the Type II solution reported here corresponds to the half-dyon configuration exhibiting negative magnetic charge and positioned along the positive $z$-axis. The total energy, magnetic charge, magnetic dipole moment, and total electric charge of both Type I and Type II solutions are calculated.

The following section offers a brief overview of the SU(2) Yang-Mills-Higgs field theory. The construction of the half-dyon solution, including the magnetic ansatz utilized and its fundamental characteristics, is outlined in section \ref{ansatz}. Subsequently, the numerical results obtained from the calculations of the half-dyon solution are presented and discussed in section \ref{results}. Finally, we conclude with some comments in section \ref{conclusions}.

\section{The SU(2) Yang-Mills-Higgs Theory}\label{SU2YMH}
The SU(2) Yang-Mills-Higgs Lagrangian in $3+1$ dimensions is given by
\begin{equation}\label{Lagrangian}
\mathcal{L}=-\frac{1}{4}F^a_{\mu\nu}F^{a\mu\nu}-\frac{1}{2}D^\mu\Phi^a D_\mu\Phi^a-\frac{1}{4}\lambda\bigl(\Phi^a\Phi^a - \nu^2\bigr)^2,
\end{equation}
with the covariant derivatives of the gauge field strength tensor and Higgs field, respectively
\begin{align}
F^a_{\mu\nu} &= \partial_\mu A^a_\nu - \partial_\nu A^a_\mu + g\epsilon^{abc}A^b_\mu A^c_\nu, \nonumber\\
D_\mu\Phi^a &= \partial_\mu\Phi^a + g\epsilon^{abc}A^b_\mu\Phi^c.
\end{align}
Here $g$ is the coupling constant of gauge field. The vacuum expectation value of the Higgs field is $\nu = \mu/\sqrt{\lambda}$
where $\mu$ is the Higgs field mass and $\lambda$ is the Higgs self-coupling constant. Greek indices refer to Minkowski spacetime and takes values $\mu, \nu = 0,1,2,3$ whereas Latin indices refer to the internal isospin and takes values $a,b,c = 1,2,3$. The metric used is $-g_{00} = g_{11} = g_{22} = g_{33} = +1 $. The equations of motion are
\begin{align}\label{EOM}
D^\mu F^a_{\mu\nu} &= \partial^\mu F^a_{\mu\nu} + g\epsilon^{abc}A^{b\mu}F^c_{\mu\nu} = g\epsilon^{abc}\Phi^b D_\nu\Phi^c, \nonumber\\
D^\mu D_\mu \Phi^a &= \lambda\Phi^a(\Phi^b\Phi^b-\nu^2) .
\end{align}
As the parameters $\mu$ and $\lambda$ approach zero, the Higgs potential approaches zero as well, and it becomes possible to obtain self-dual solutions by solving the first-order partial differential Bogomol’nyi equation,
\begin{equation}
B^a_i \pm D_i\Phi^a = 0,~~ \text{where}~~B^a_i = -\frac{1}{2}\epsilon_{ijk}F^a_{jk}.
\end{equation}
The electromagnetic field tensor proposed by 't Hooft \cite{kn:2} upon symmetry breaking is
\begin{align}\label{thooftdef}
F_{\mu\nu} &= \hat{\Phi}^a F^a_{\mu\nu} - \frac{1}{g}\epsilon^{abc}\hat{\Phi}^a D_\mu\hat{\Phi}^b D_\nu\hat{\Phi}^c\nonumber\\
&= \partial_\mu A_\nu - \partial_\nu A_\mu -\frac{1}{g}\epsilon^{abc}\hat{\Phi}^a\partial_\mu\hat{\Phi}^b\partial_\nu\hat{\Phi}^c,
\end{align}
where $A_\mu = \hat{\Phi}^a A^a_\mu$, $\hat{\Phi}^a = \Phi^a / |\Phi|$, and $|\Phi| = \sqrt{\Phi^a\Phi^a}$. We can view the 't Hooft electromagnetic field tensor to be composed of both the gauge part $G_{\mu\nu}$ and the Higgs part $H_{\mu\nu}$ of the electromagnetic field tensor
\begin{equation}
G_{\mu\nu} = \partial_\mu A_\nu - \partial_\nu A_\mu ,~~ H_{\mu\nu} = -\frac{1}{g}\epsilon^{abc}\hat{\Phi}^a\partial_\mu\hat{\Phi}^b\partial_\nu\hat{\Phi}^c,
\end{equation}
subsequently we can define the magnetic field in a similar fashion
\begin{equation}\label{thooftmag}
B_i = -\frac{1}{2}\epsilon_{ijk}F_{jk} = B^G_i + B^H_i ,
\end{equation}
where $B^G_i$ and $B^H_i$ are the gauge part and Higgs part of the magnetic field respectively. The net magnetic charge of the system is
\begin{equation}\label{magneticcalc}
M = \int \partial^i B_i ~d^3 x = \oint d^2\sigma_i B_i.
\end{equation}

\section{The Axially Symmetric Half-Dyon}\label{ansatz}
\subsection{Ansatz}
We consider the following time independent axially symmetric magnetic ansatz that leads to the half-dyon solutions
\begin{align}\label{dyonansatz}
gA^a_i &= -\frac{1}{r}\psi_1(r,\theta)\hat{\phi}^a\hat{\theta}_i + \frac{n}{r\sin\theta}P_1(r,\theta)\hat{\theta}^a\hat{\phi}_i + \frac{1}{r}R_1(r,\theta)\hat{\phi}^a \hat{r}_i - \frac{n}{r\sin\theta}P_2(r,\theta)\hat{r}^a\hat{\phi}_i, \nonumber\\
gA^a_0 &= \tau_1(r,\theta)\hat{r}^a + \tau_2(r,\theta)\hat{\theta}^a,\nonumber  \\
g\Phi^a &= \Phi_1(r,\theta)\hat{r}^a + \Phi_2(r,\theta)\hat{\theta}^a.
\end{align}
The spatial spherical coordinate orthonormal unit vectors are
\begin{align}
\hat{r}_i &= \sin\theta\cos \phi \,\delta_{i1} + \sin\theta\sin \phi\,\delta_{i2} + \cos\theta\,\delta_{i3},\nonumber\\
\hat{\theta}_i &= \cos\theta\cos \phi \,\delta_{i1} + \cos\theta\sin \phi\,\delta_{i2} - \sin\theta\,\delta_{i3}, \nonumber\\
\hat{\phi}_i &= -\sin\phi \,\delta_{i1} + \cos\phi\,\delta_{i2},
\end{align}
and the isospin coordinate orthonormal unit vectors are
\begin{align}
\hat{r}^a &= \sin\theta\cos n\phi \,\delta^a_1 + \sin\theta\sin n\phi\,\delta^a_2 + \cos\theta\,\delta^a_3,\nonumber\\
\hat{\theta}^a &= \cos\theta\cos n\phi \,\delta^a_1 + \cos\theta\sin n\phi\,\delta^a_2 - \sin\theta\,\delta^a_3, \nonumber\\
\hat{\phi}^a &= -\sin n\phi \,\delta^a_1 + \cos n\phi\,\delta^a_2.
\end{align}
The $\phi$-winding number, $n$, is a natural number corresponds to the topological charge of dyon soluton.
\subsection{The Higgs Field}
The general Higgs fields in the spherical and rectangular coordinate systems are 
\begin{align}
g\Phi^a &= \Phi_1(x)\hat{r}^a + \Phi_2(x)\hat{\theta}^a + \Phi_3(x)\hat{\phi}^a\nonumber\\
&= \tilde{\Phi}_1(x)\delta^{a1} + \tilde{\Phi}_2\delta^{a2} + \tilde{\Phi}_3\delta^{a3},
\end{align}
respectively, where
\begin{align}
\tilde{\Phi}_1 &= \sin\theta\cos n\phi\,\Phi_1 + \cos\theta\cos n\phi\,\Phi_2 - \sin n\phi\,\Phi_3 = |\Phi|\sin\alpha\cos\beta, \nonumber\\
\tilde{\Phi}_2 &= \sin\theta\sin n\phi\,\Phi_1 + \cos\theta\sin n\phi\,\Phi_2 + \cos n\phi\,\Phi_3 = |\Phi|\sin\alpha\sin\beta,\nonumber \\
\tilde{\Phi}_3 &= \cos\theta \,\Phi_1 - \sin\theta\,\Phi_2 = |\Phi|\cos\alpha .
\end{align}
The axially symmetric Higgs unit vectr in the rectangular coordinate system is given by
\begin{align}\label{cosalpha}
&\hat{\Phi}^a = \sin\alpha\cos\gamma \,\delta^{a1} + \sin\alpha\sin\gamma \,\delta^{a2} + \cos\alpha\,\delta^{a3},\nonumber\\
&\cos\alpha = h_1(r,\theta)\cos\theta - h_2(r,\theta)\sin\theta,~~ \sin\alpha= h_1(r,\theta)\sin\theta+h_2(r,\theta)\cos\theta, \nonumber \\
&h_1(r,\theta) = \frac{\Phi_1}{|\Phi|},~~ h_2(r,\theta) = \frac{\Phi_2}{|\Phi|},~~ \gamma = n\phi. 
\end{align}
\subsection{The Magnetic and Electric Fields}
Following the definition of $\cos\alpha$ (\ref{cosalpha}), we can rewrite the Higgs part of the 't Hooft magnetic field (\ref{thooftmag}) in a different form
\begin{equation}\label{higgsmag}
g B^H_i = -n\epsilon_{ijk} \partial^j \cos\alpha \,\partial^k\phi .
\end{equation}
The gauge part of the magnetic field can similarly be rewritten as
\begin{equation}\label{gaugemag}
g B^G_i = -n\epsilon_{ijk}\partial^j\cos\kappa\, \partial^k\phi ,
\end{equation}
where $\cos\kappa = P_1 h_2 - P_2 h_1$. Hence the sum of the Higgs part (\ref{higgsmag}) and gauge part (\ref{gaugemag}) gives a different definition of 't Hooft magnetic field
\begin{equation}\label{thooftgauge}
gB_i = -n\epsilon_{ijk}\partial_j(\cos\alpha+\cos\kappa)\partial_k\phi = -n\epsilon_{ijk}\partial_j\mathcal{A}_k,
\end{equation}
where $\mathcal{A}_i$ is the 't Hooft gauge potential. The magnetic field lines of the half-dyon system can be pictured by representing the contour lines of $(\cos\alpha + \cos\kappa) = $ constant on the vertical plane $\phi = 0$.

For the electric charge, the Abelian electric field is
\begin{equation}
\mathcal{E}_i = F_{i0} = \partial_i A_0 = \partial_i\{\tau_1\cos(\alpha-\theta)+\tau_2\sin(\alpha-\theta)\}=\partial_i|\tau|,
\end{equation}
where $|\tau| = \sqrt{\tau_1^2+\tau_2^2}$, since the time component of the gauge field $A_0^a$ is parallel to the Higgs field $\Phi^a$ in isospin space. In contrast to the magnetic field, the electric field changes in proportion to the electric charge parameter, $0\leq \eta <1$. By setting $\eta = 0$, it is possible to turn off the electric field. We can calculate the total electric charge $Q$ of the system by numerically evaluating the volume integration and consider the case of axially symmetric dyon
\begin{equation}
Q = \int_V \partial^i \mathcal{E}_i d^3x = 2\pi \iint \partial^i \mathcal{E}_i r^2\sin\theta\,dr\,d\theta.
\end{equation}

At spatial infinity in the Higgs vacuum, the non-Abelian components of the gauge potential become zero, and the non-Abelian electromagnetic field approaches a particular value, which is given by
\begin{align}
F^a_{\mu\nu}\big|_{r\rightarrow\infty} &= \bigl[\partial_\mu A_\nu - \partial_\nu A_\mu - \frac{1}{g}\epsilon^{bcd}\hat{\Phi}^b\partial_\mu\hat{\Phi}^c\partial_\nu\hat{\Phi}^d\bigr]\hat{\Phi}^a\nonumber\\
&= F_{\mu\nu}\hat{\Phi}^a,
\end{align}
where $F_{\mu\nu}$ is the 't Hooft electromagnetic field. However, Coleman \cite{kn:29} argues that there is no single unique way to represent the Abelian electromagnetic field in the area surrounding the monopole outside the Higgs vacuum at finite values of $r$. One approach, proposed by 't Hooft, as in Eq.~(\ref{thooftdef}), yields a magnetic field with the unique feature that the magnetic charge density vanishes when $|\Phi| \neq 0$, whereas $\partial^i B_i \neq 0$ when $|\Phi| = 0$, and the magnetic charges are located at these points. Consequently, according to 't Hooft's definition of the electromagnetic field, the magnetic charges are individual located at the point zeros of the Higgs field. This leads to a discrete magnetic charge at a specific point, and there is no magnetic charge distribution throughout the space. Another proposal for the Abelian electromagnetic field was offered by Bogomol’nyi \cite{kn:4} and Faddeev\cite{kn:30,kn:31}, in which the magnetic and electric fields are less singular, is expressed respectively as follows:
\begin{equation}\label{Faddeevdef}
\mathcal{B}_i = B^a_i \Bigl(\frac{\Phi^a}{\nu}\Bigr),~~~~ \mathcal{E}_i = E^a_i \Bigl(\frac{\Phi^a}{\nu}\Bigr),
\end{equation}
where $\nu$ is the vacuum expectation value of the Higgs field. With this definition, we can obtain the electromagnetic field in a less singular form and be able to visualize both electric and magnetic charge distributions in finite $r$ region produced by the non-Abelian components of the gauge field. At spatial infinity in Higgs vacuum, both definitions of the electromagnetic field (\ref{thooftdef}) and (\ref{Faddeevdef}) become identical. To obtain the 3D surface graph of magnetic charge density, we will instead plot using the weighted magnetic charge density which is defined as $M_d = (\partial^i\mathcal{B}_i) r^2\sin\theta$ because the nature of magnetic charge density of half-dyon solution is singular yet integrable. Similar technique will be used when plotting the weighted electric charge density, $Q_d = (\partial^i \mathcal{E}_i) r^2\sin\theta$.

\subsection{The Magnetic Dipole Moment and Angular Momentum}
From Maxwell's electromagnetic theory, the 't Hooft gauge potential, $\mathcal{A}_i$, Eq.~(\ref{thooftgauge}) at large $r$ tends to
\begin{align}
&\mathcal{A}_i = (\cos\alpha+\cos\kappa)\partial_i\phi\big|_{r\rightarrow\infty} = \frac{\hat{\phi}_i}{r\sin\theta}\bigg\{\frac{n}{2}(\cos\theta + 1)+\frac{F_G(\theta)}{r}\bigg\},\nonumber \\
&F_G(\theta) = \mu_m \sin^2\theta = nr\bigl[(-P_2+\cos\theta)h_1 + (P_1-\sin\theta)h_2 - \frac{1}{2}(\cos\theta+1)\bigr]\big|_{r\rightarrow\infty},
\end{align}
where $\mu_m$ is the dimensionless magnetic dipole moment of the half-monopole. We can obtain the value of magnetic dipole moment of the half-monopole numerically by plotting $F_G(\theta)$ versus angle $\theta$. The value of $\mu_m$ can then be read directly from the graph of $F_G(\theta)$ versus $\theta$ at $\theta = \pi/2$.

The energy momentum of the SU(2) YMH theory is given by
\begin{equation}
\theta_{\mu\nu} = F^{a\beta}_\mu F^a_{\mu\nu} - \frac{1}{4}g_{\mu\nu}F^a_{\alpha\beta}F^{a\alpha\beta} + D_\mu\Phi^a D_\nu\Phi^a - \frac{1}{2}\bigl( D_\alpha\Phi^a D^\alpha\Phi^a+\frac{1}{4}\lambda(\Phi^a\Phi^a-\nu^2)^2\bigr).
\end{equation}
With the magnetic ansatz (\ref{ansatz}), the Poynting vector of the system can be written as~\cite{kn:32,kn:33}
\begin{align}
\theta_{0i} &= F^{aj}_0F^{a}_{ij} + D_0\Phi^a D_i\Phi^a \nonumber\\
&= [ -F^{aj}_0 (F^a_{jk}\hat{\phi}_k) + D_0\Phi^a (D_k\Phi^a \hat{\phi}_k) ] \hat{\phi}_i  \nonumber\\
&= -\frac{1}{r\sin\theta} \hat{\phi}_i \partial^j (F^a_{0j}W^a r\sin\theta),
\end{align}
where $W^a = A^a_i\hat{\phi}_i - \Psi^a$ and $\Psi^a = (-n\cot\theta  \hat{r}^a + n \hat{\theta}^a) /r $. For the half-dyon solutions, the $W^a$ is
\begin{equation}
W^a = \frac{n}{r} \Bigl(-\frac{P_2 (r,\theta)}{\sin\theta} + \cot\theta \Bigr) \hat{r}^a + \frac{n}{r} \Bigl( \frac{P_1(r,\theta)}{\sin\theta} - 1 \Bigr) \hat{\theta}^a.
\end{equation}
The angular momentum density along the $z$-axis is given by
\begin{equation}
 j_z = \epsilon_{kij} \hat{\rho}_i \rho \theta_{0j}\delta^3_k = -\partial^i(F^a_{0i}W^a r\sin\theta),
\end{equation}
where $\rho = \sqrt{x^2_1+x^2_2}=r\sin\theta$ and $\hat{\rho}_i = \cos\phi\delta_{i1} + \sin\phi\delta_{i2}$. Hence the total angular momentum is
\begin{align}\label{eqangular}
J_z &= - \int d^3 x \partial^i(F^a_{0i}W^a r\sin\theta) \nonumber\\
&= - 2\pi \int^\pi_0 (F^a_{0i}W^a r \sin\theta)\hat{r}_i r^2\sin\theta d\theta \Big| _{r\rightarrow\infty} \nonumber\\
&= 2\pi \int^\pi_0 r^2\sin\theta (1+\cos\theta)\Bigl(\partial_r\tau_1(r,\theta)\cos\frac{\theta}{2} -\partial_r\tau_2(r,\theta)\sin\frac{\theta}{2}\Bigr) d\theta \Big| _{r\rightarrow\infty} \nonumber \\
&= 2\pi \lim_{r\rightarrow\infty} r^2\partial_r\tau(r),
\end{align}
if we assume that the time component of the gauge field, $A^a_0$, is parallel to the Higgs field, $\Phi^a$, at spatial infinity. Hence $\tau_1(r,\theta) \rightarrow \tau(r) \cos\theta/2$ and $\tau_2(r,\theta) \rightarrow \tau(r) \sin\theta/2$ as $r\rightarrow\infty$. Thus from Eq.~(\ref{eqangular}), $J_z = Q/2$ and the half-dyon solutions possess kinetic energy of rotation.

\subsection{The Energy}
The total energy of the system is
\begin{equation}
E = \int_V \varepsilon\, d^3 x ,
\end{equation}
where $\varepsilon$ is the energy density of the SU(2) Yang-Mills-Higgs theory given by
\begin{equation}
\varepsilon = B^a_i B^a_i + E^a_i E^a_i + D_i \Phi^a D_i \Phi^a + D_0\Phi^a D_0\Phi^a + \frac{\lambda}{2}(\Phi^a\Phi^a - \nu^2)^2.
\end{equation}
By dimensionless transformations, we can change $\varepsilon$ to a dimensionless value $\tilde{\varepsilon}$. Consider the case with axial symmetry, we have
\begin{equation}
E = 2\pi \iint \varepsilon r^2\sin\theta\,dr\,d\theta =\frac{2\pi\nu}{g} \iint \tilde{\varepsilon} x^2\sin\theta\,dx\,d\theta.
\end{equation}
Hence the total energy of the system $E$ is in unit of $2\pi\nu/g$.

\subsection{The Boundary Conditions and Numerical Calculations}
The numerical half-dyon solution is solved by substituting the Ansatz (\ref{dyonansatz}) into the equations of motion (\ref{EOM}) which resulted in eight coupled nonlinear second order partial differential equations. By applying different sets of boundary conditions, we can obtain half-dyon configurations with positive (Type I) or negative (Type II) magnetic charge. For Type I solution, the half-dyon solution at large distances ($r \rightarrow \infty$) is given by \cite{kn:14,kn:15}
\begin{align} \label{boundary1}
\psi_1 &= \frac{1}{2},~~ P_1 = \sin\theta - \frac{1}{2}(1+\cos\theta)\sin\Bigl(\frac{\theta}{2}\Bigr),\nonumber\\
R_1 &= 0,~~ P_2 = \cos\theta - \frac{1}{2}(1+\cos\theta)\cos\Bigl(\frac{\theta}{2}\Bigr),\nonumber\\
\Phi_1 &= \cos\Bigl(\frac{\theta}{2}\Bigr),~~ \Phi_2 =-\sin\Bigl(\frac{\theta}{2}\Bigr),\nonumber\\
\tau_1 &= \eta\cos\Bigl(\frac{\theta}{2}\Bigr),~~ \tau_2 =-\eta\sin\Bigl(\frac{\theta}{2}\Bigr).
\end{align}
In this region, it is obvious that $\Phi_1 \propto \tau_1$ and $\Phi_2 \propto \tau_2$ that is the time component of gauge field and the Higgs field are parallel in the isospin space. We have the trivial vacuum solution close to the origin $(r = 0)$. The asymptotic solution and boundary conditions at $r=0$ which will produce finite energy half-dyon solution are 
\begin{align} \label{boundary2}
\psi_1(0,\theta) = P_1(0,\theta) = R_1(0,\theta) = P_2(0,\theta) &= 0,\nonumber\\
\sin\theta\,\Phi_1(0,\theta) + \cos\theta\,\Phi_2(0,\theta) &= 0,\nonumber\\
\sin\theta\,\tau_1(0,\theta) + \cos\theta\,\tau_2(0,\theta) &= 0,\nonumber\\
\frac{\partial}{\partial r}\{\cos\theta\,\Phi_1(r,\theta) - \sin\theta\,\Phi_2(r,\theta)\}\Big|_{r=0} &= 0,\nonumber\\
\frac{\partial}{\partial r}\{\cos\theta\,\tau_1(r,\theta) - \sin\theta\,\tau_2(r,\theta)\}\Big|_{r=0} &= 0.
\end{align}
The boundary conditions will be imposed on the profile functions along the positive $z$-axis $(\theta = 0)$ as follows:
\begin{equation} \label{boundary3}
\partial_\theta\psi_1 = P_1 = R_1 = P_2 = \partial_\theta\Phi_1 = \Phi_2 = \partial_\theta\tau_1 = \tau_2 = 0.
\end{equation}
The boundary conditions imposed along the negative $z$-axis ($\theta = \pi$) are
\begin{equation} \label{boundary4}
\partial_\theta\psi_1 = P_1 = R_1 = \partial_\theta P_2 = \Phi_1 = \partial_\theta\Phi_2 = \tau_1 = \partial_\theta\tau_2 = 0.
\end{equation}

To obtain the Type II solution, the following boundary conditions are used at $r \rightarrow \infty$,
\begin{align}\label{boundary1-negative}
\psi_1 &= \frac{1}{2},~~ P_1 = \sin\theta - \frac{1}{2}(1-\cos\theta)\cos\Bigl(\frac{\theta}{2}\Bigr),\nonumber\\
R_1 &= 0,~~ P_2 = \cos\theta + \frac{1}{2}(1-\cos\theta)\sin\Bigl(\frac{\theta}{2}\Bigr),\nonumber\\
\Phi_1 &= -\sin\Bigl(\frac{\theta}{2}\Bigr),~~ \Phi_2 =-\cos\Bigl(\frac{\theta}{2}\Bigr),\nonumber\\
\tau_1 &= -\eta\sin\Bigl(\frac{\theta}{2}\Bigr),~~ \tau_2 =-\eta\cos\Bigl(\frac{\theta}{2}\Bigr).
\end{align}
At the origin ($r=0$), the same boundary conditions (\ref{boundary2}) are used. Along the positive $z$-axis ($\theta=0$), the boundary conditions are
\begin{equation}\label{boundary3-negative}
\partial_\theta\psi_1 = P_1 = R_1 = \partial_\theta P_2 = \Phi_1 = \partial_\theta\Phi_2 = \tau_1 = \partial_\theta\tau_2 = 0,
\end{equation}
and along the negative $z$-axis ($\theta$=$\pi$), the boundary conditions are
\begin{equation}\label{boundary4-negative}
\partial_\theta\psi_1 = P_1 = R_1 = P_2 = \partial_\theta\Phi_1 = \Phi_2 = \partial_\theta\tau_1 = \tau_2 = 0.
\end{equation}

For the purpose of performing numerical calculations, we transform the spatial coordinates $r$ into dimensionless coordinates $x$ using the relationship $x = g\nu r$, and rescale the profile functions by factors of $\nu$ and $g\nu$: $\Phi_1 \rightarrow \nu\Phi_1$, $\Phi_2 \rightarrow \nu\Phi_2$, $\tau_1 \rightarrow g\nu\tau_1$, and $\tau_2 \rightarrow g\nu\tau_2$. This process results in a system of equations that depends solely on $n$, $\eta$, and $\beta$, where $\beta^2 = \lambda/g^2$. Furthermore, it eliminates any dependence on the original variables $g$, $\nu$, and $\lambda$. We discretize the system of equations using finite difference approximation on a non-equidistant grid of size $70 \times 60$. The grid covers the integration regions of $0\leq \tilde{x} \leq 1 $ and $ 0 \leq \theta \leq \pi$, where $\tilde{x} = x/(x+1)$ is the compactified coordinate. The derivatives with respect to $x$ are then replaced with $\tilde{x}$ using the relations, $\partial_x \rightarrow (1-\tilde{x})^2\partial_{\tilde{x}}$ and $\partial_{xx} \rightarrow (1-\tilde{x})^4\partial_{\tilde{x}\tilde{x}}-2(1-\tilde{x})^3\partial_{\tilde{x}}$. We solve the equations of motion Eq.~\ref{EOM} for $\phi$-winding number, $1\leq n \leq 4$, the electric charge parameter, $0 \leq\eta <\eta_{max}$, and with the Higgs self-coupling constant, $0 \leq \beta \leq 1$. The error arises from our numerical results is $\mathcal{O}(10^{-3})$.

\section{Results}\label{results}
By implementing the appropriate boundary conditions for Type I solution, Eqs.~(\ref{boundary1}) to (\ref{boundary4}), we compute numerical solutions for the profile functions $\psi_1$, $R_1$, $P_1$, $P_2$, $\tau_1$, $\tau_2$, $\Phi_1$, and $\Phi_2$ for $\phi$-winding number ranging $ 0 \leq n \leq 4$, electric charge parameters ranging $ 0 \leq \eta < \eta_{max}$ and Higgs self-coupling constant $ 0 \leq \beta \leq 1$. It is important to note that no solutions were found for $n > 4$. For $n = 1,2$ and 3, $\eta_{max} = 1$ whereas for $n=4$, $\eta_{max} = 0.9354$. The presence of electric charge is directly related to the value of $\eta$, which means that we have the standard half-monopole solutions when $\eta=0$, and dyon solutions can be obtained when $\eta > 0$.

\begin{figure}[b]
\begin{center}
	\includegraphics[width=\textwidth,keepaspectratio]{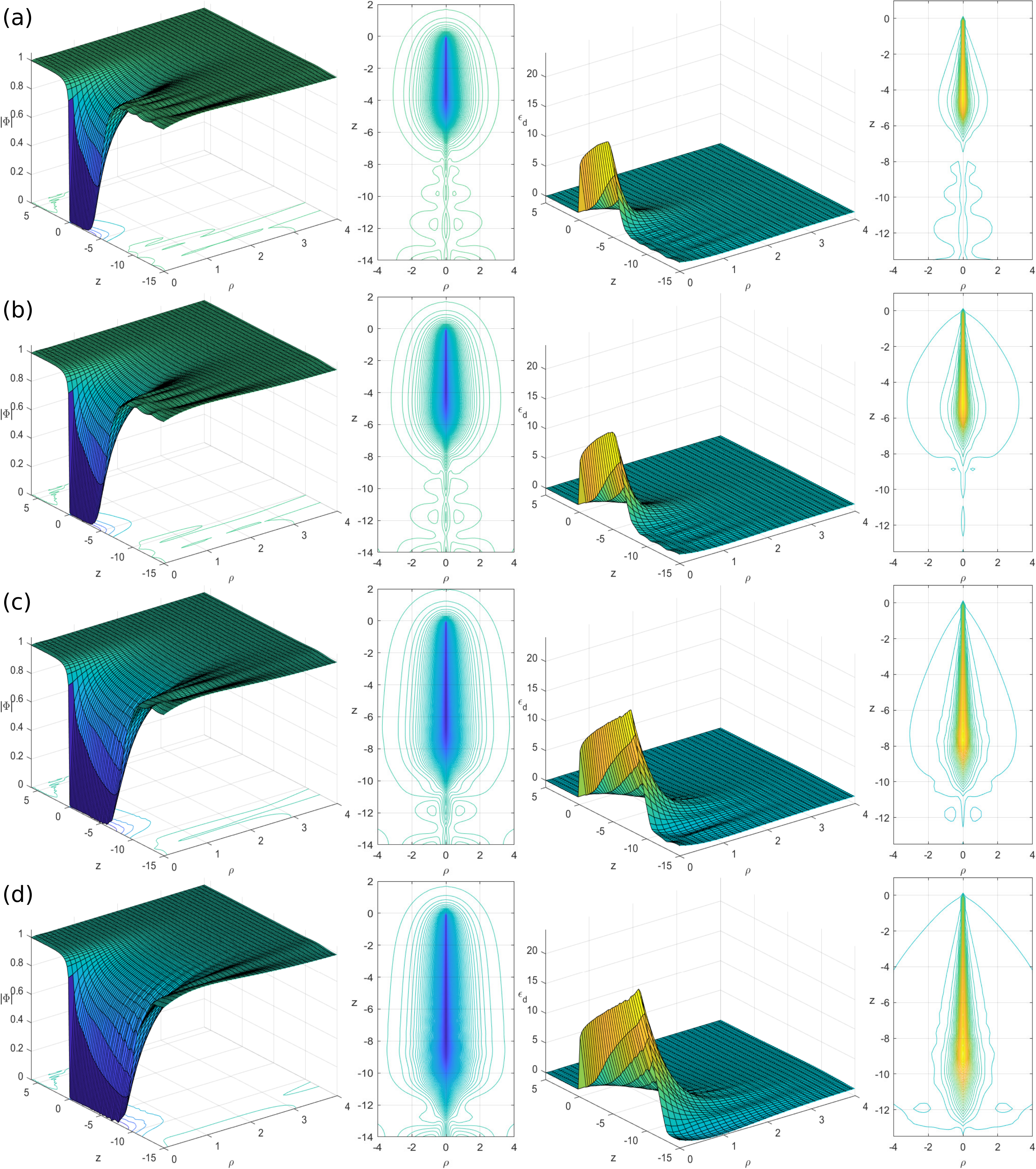}
\end{center}
\caption{Higgs modulus (left) and weighted energy density (right) of Type I solution at $\beta=0.7782$ with their inset contour for (a) $n=2$, $\eta=0.6$; (b) $n=2$, $\eta=0.9$; (c) $n=3$, $\eta=0.9$; (d) $n=4$, $\eta=0.9$.}\label{higgsenergyplot} 
\end{figure}

The general shape of the dyons resembles the findings of Teh et al. \cite{kn:14,kn:15}, which is a balloon-like shape that expands from the origin and extends along the negative $z$-axis. Fig.~\ref{higgsenergyplot} illustrates the 3D and contour plots of the Higgs modulus, $|\Phi|$. From the graphs of Higgs modulus, we observe that the inverted cone shape stretches further down the negative $z$-axis as $\eta$ increases, and the zeros of Higgs modulus also increase along the negative $z$-axis. Similarly, the length of the Higgs modulus along the negative $z$-axis also increases, but at a greater rate as $n$ increases than with increases in $\eta$.   The peak value of the weighted energy density, $\epsilon_d$, is 15.15 at $z=-4.916$ ($n=2$, $\eta=0.9$) and 22.09 at $z=-8.849$ ($n=4$, $\eta=0.9$). The energy is integrable along the negative $z$-axis indicating that the half-dyon solutions possess finite energy.

\begin{figure}[b]
\begin{center}
	\includegraphics[width=\textwidth,keepaspectratio]{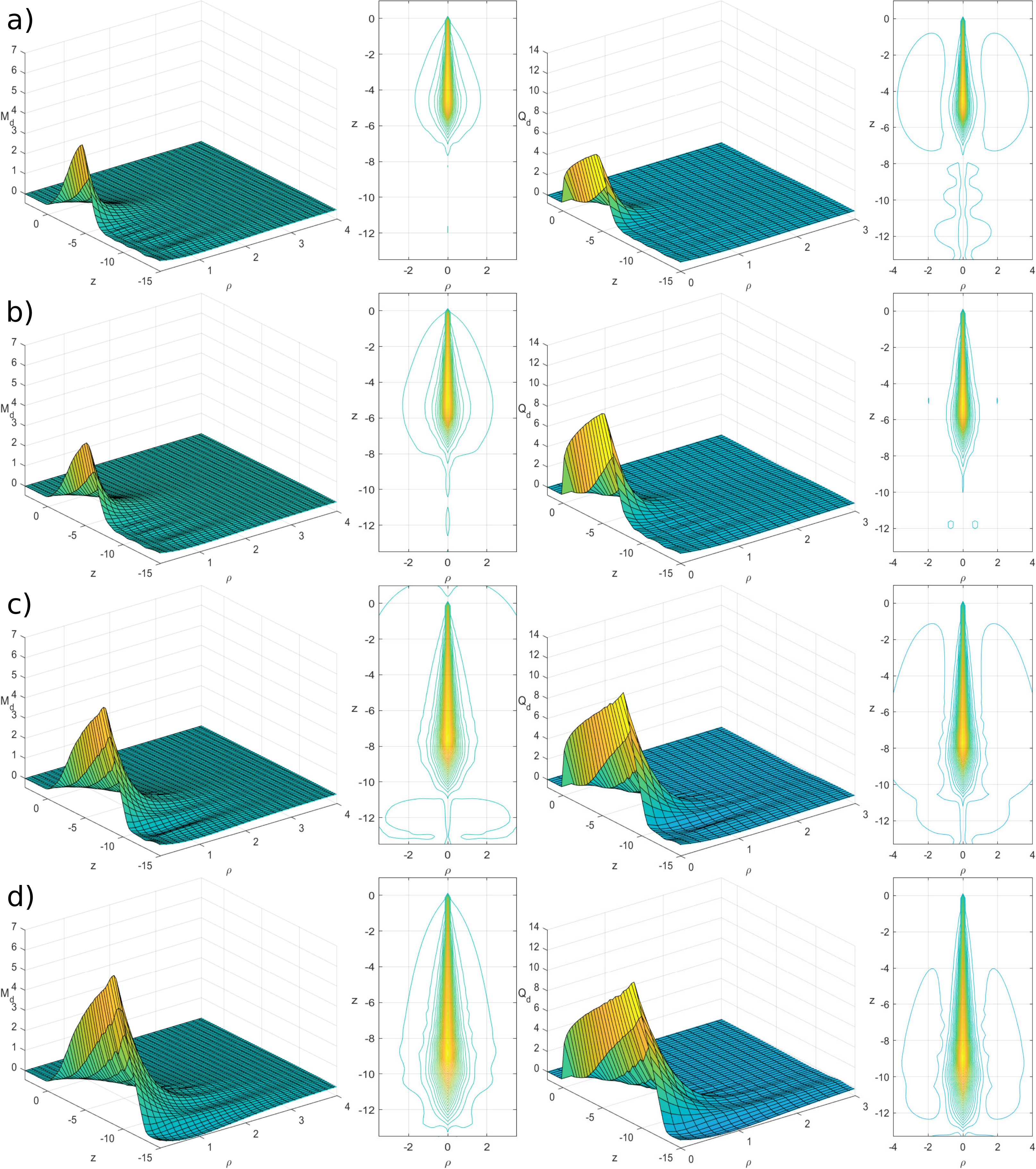}
\end{center}
\caption{Weighted magnetic charge density (left) and weighted electric charge density (right) of Type I solution at $\beta=0.7782$ with their inset contour for (a) $n=2$, $\eta=0.6$; (b) $n=2$, $\eta=0.9$; (c) $n=3$, $\eta=0.9$; (d) $n=4$, $\eta=0.9$.}\label{electplot} 
\end{figure}

Fig.~\ref{electplot} depicts the plots of the weighted magnetic charge density, $M_d$ and weighted electric charge density, $Q_d$, of the half-dyon system. As $n$ and $\eta$ increase, the peak of the electric charge density extends along the negative $z$-axis from the origin. This observation is confirmed by the contour plots, which demonstrate that the expansion is restricted solely to the negative $z$ direction. Additionally, we note that the total electric charge $Q$ of the system increases as the magnitude of the electric charge density rises in conjunction with an increase in $n$.

From Fig.~\ref{electplot}, the behavior of the weighted magnetic charge density, $M_d$, is similar to that of the weighted electric charge density. Specifically, as $n$ and $\eta$ increase, the size of the distribution only changes in the negative $z$-axis direction. We notice that while an increase in $\eta$ does not significantly affect the magnetic field strength, it elongates the magnetic charge distribution along the negative $z$-axis. This observation is supported by the corresponding contour plots.

\begin{figure}[b]
\begin{center}
	\includegraphics[width=\textwidth,keepaspectratio]{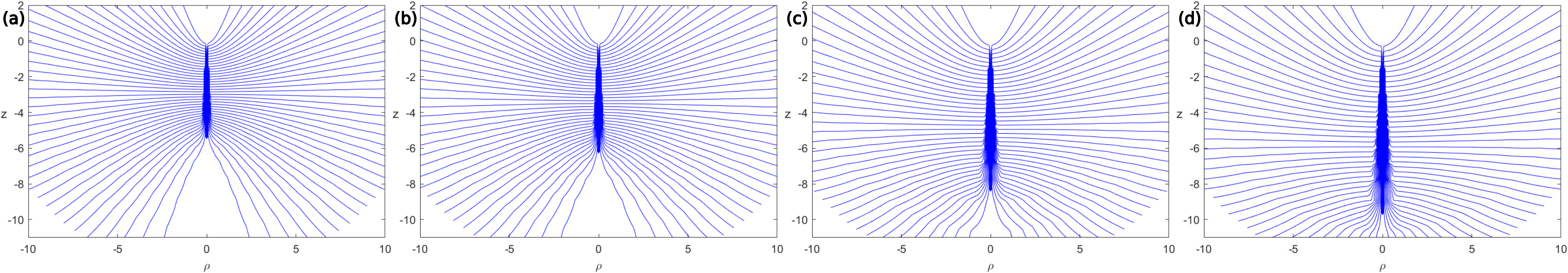}
\end{center}
\caption{Magnetic field lines of Type I solution at $\beta=0.7782$ for (a) $n=2$, $\eta=0.6$; (b) $n=2$, $\eta=0.9$; (c) $n=3$, $\eta=0.9$; (d) $n=4$, $\eta=0.9$.}\label{magneticplot} 
\end{figure}

Fig.~\ref{magneticplot} outlines the magnetic field lines of the half-dyon configuration. As $\eta$ increases, the magnetic charge extends gradually towards the negative $z$-axis, with this extension becoming more pronounced as $n$ increases. Thus, we can conclude that the half-dyon system possesses both magnetic and electric charge densities that extend from the origin toward the negative $z$-axis. We also calculate the magnetic charge of the system with half-dyon numerically using Eq.~(\ref{magneticcalc}) and find its net magnetic charge is $2n\pi/g$ for $1\leq n \leq 4$.

\begin{figure}
\begin{center}
\includegraphics[width=\textwidth,keepaspectratio]{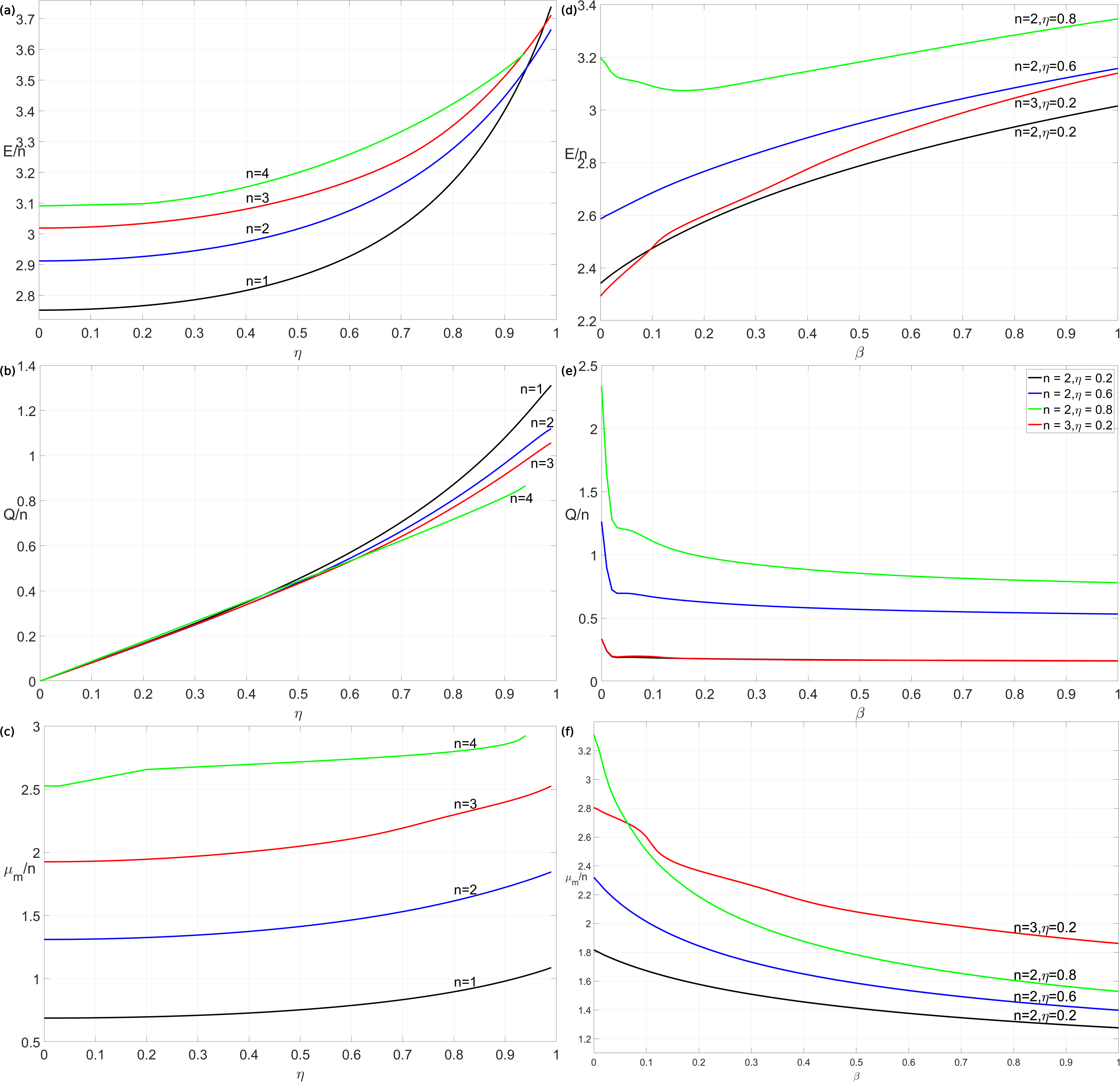}
\end{center}

\caption{Plots of (a) total energy per $n$, $E/n$, (b) total electric charge per $n$, $Q/n$, and (c) magnetic dipole moment per $n$, $\mu_m/n$ versus $\eta$ at $\beta = 0.7782$. Plots of (d) $E/n$, (e) $Q/n$, and (f) $\mu_m/n$ versus $\beta$ for $n=2, \eta=0.2,0.6,0.8$ and $n=3, \eta=0.2$. }\label{lineplot}
\end{figure}

The numerical values of the total electric charge per $n$, $Q/n$, magnetic dipole moment per $n$, $\mu_m/n$, and total energy per $n$, $E/n$, for the half-dyon system are computed and presented in Table~\ref{table1}. The graphs in Fig.~\ref{lineplot}(a)-(c) illustrate the relationship between these values and $\eta$. The graphs indicate that $Q/n$ decreases with increasing $n$, while $\mu_m/n$ increases monotonically as $n$ and $\eta$ increase. For $E/n$ with $n>1$ there exists a repulsive phase. For $\eta > 0.94 $, the energy per $n$ of $n=2$ case is lower than the $n=1$ case which implies that the $n=2$ configuration is more stable in this region. 

\begin{table}[t]
\centering
\caption{Selected values for total electric charge per $n$, $Q/n$, in unit of $2\pi/g$, magnetic dipole moment per $n$, $\mu_m/n$, in unit of $1/g$ and total energy per $n$, $E/n$ in unit of $2\pi\nu/g$ with Higgs self-coupling constant $\beta = 0.7782$ for $\phi$-winding number $1\leq n \leq 4$.} 
{\begin{tabular}{@{}ccccccccccc@{}}
\hline
$n = 1$ \\
\hline
$\eta$ & 0 & 0.2 & 0.3 & 0.4 & 0.5 & 0.6 & 0.7 & 0.8 & 0.9 & 0.99\\ 
\hline
$Q/n$ & 0  & 0.1671 & 0.2552 & 0.3495 & 0.4529 & 0.5695 & 0.7057 & 0.8706 & 1.0790 & 1.3117 \\
$\mu_m/n$ & 0.6879  & 0.6974 & 0.7097 & 0.7280 & 0.7534 & 0.7878 & 0.8339 & 0.8960 & 0.9809 & 1.0873\\
$E/n$ & 2.7519 & 2.7661 & 2.7855 & 2.8157 & 2.8606 & 2.9266 & 3.0241 & 3.1711 & 3.4013 & 3.7394\\
\hline
\\
\hline
$n = 2$ \\
\hline
$\eta$ & 0 & 0.2 & 0.3 & 0.4 & 0.5 & 0.6 & 0.7 & 0.8 & 0.9 & 0.99\\ 
\hline
$Q/n$ & 0  & 0.1633 & 0.2488 & 0.3392 & 0.4367 & 0.5442 & 0.6652 & 0.8039 & 0.9653 & 1.1199 \\
$\mu_m/n$ & 1.3103  & 1.3256 & 1.3452 & 1.3740 & 1.4131 & 1.4644 & 1.5307 & 1.6153 & 1.7228 & 1.8453\\
$E/n$ & 2.9120 & 2.9261 & 2.9449 & 2.9739 & 3.0159 & 3.0755 & 3.1592 & 3.2773 & 3.4465  & 3.6652\\ 
\hline
\\
\hline
$n = 3$ \\
\hline
$\eta$ & 0 & 0.2 & 0.3 & 0.4 & 0.5 & 0.6 & 0.7 & 0.8 & 0.9 & 0.99\\
\hline
$Q/n$ & 0  & 0.1648 & 0.2503 & 0.3390 & 0.4317 & 0.5305 & 0.6413 & 0.7700 & 0.9147 & 1.0564 \\
$\mu_m/n$ & 1.9252  & 1.9452 & 1.9701 & 2.0046 & 2.0489 & 2.1073 & 2.1920 & 2.2969 & 2.3982 & 2.5252\\
$E/n$ & 3.0189  & 3.0335 & 3.0524 & 3.0805 & 3.1194 & 3.1717 & 3.2432 & 3.3521 & 3.5121 & 3.7119\\ 
\hline
\\
\hline
$n = 4$ \\
\hline
$\eta$ & 0 & 0.2 & 0.3 & 0.4 & 0.5 & 0.6 & 0.7 & 0.8 & 0.9 & 0.9354\\
\hline
$Q/n$ & 0  & 0.1756 & 0.2641 & 0.3530 & 0.4423 & 0.5323 & 0.6236 & 0.7172 & 0.8164 & 0.8662 \\
$\mu_m/n$ & 2.5252  & 2.6559 & 2.6763 & 2.6961 & 2.7165 & 2.7385 & 2.7643 & 2.7981 & 2.8559 & 2.9242\\
$E/n$ & 3.0911  & 3.0978 & 3.1187 & 3.1524 & 3.1988 & 3.2586 & 3.3327 & 3.4230 & 3.5347 & 3.5886\\
\hline
\end{tabular}\label{table1}}
\end{table}

Fig.~\ref{lineplot}(d)-(f) show the relationships between $E/n$, $Q/n$ and $\mu_m/n$ with varying $\beta$. We observe that $E/n$ exhibits a monotonous increase for $\eta < 0.6$ with increasing $\beta$. When comparing $n=2$ and $n=2$ at $\eta = 0.2$, the $E/n$  of $n=3$ is lower than that of $n=2$ for $\beta < 0.1$. However, beyond $\beta = 0.1$, the energy of $n=3$ surpasses that of $n=2$. For $n=2$, $\eta = 0.8$, the energy initially decreases and rises only after passing a certain value of $\beta$.

As the value of electric charge is solely dependent on the electric charge parameter $\eta$, $Q/n$ for both $n=2$ and $n=3$ when $\eta=0.2$ have very close values as in Fig~\ref{lineplot}(e). We note that the energy increases with increasing $\beta$ but both the magnetic moment and electric charge decrease while $\beta$ increases. In general, we can conclude for a fixed $n$, $E(\eta_1) > E(\eta_2)$, $Q(\eta_1) > Q(\eta_2)$ and $\mu_m (\eta_1) > \mu_m (\eta_2)$ for $\eta_1 > \eta_2$.

\begin{figure}
\begin{center}
\includegraphics[width=\textwidth,keepaspectratio]{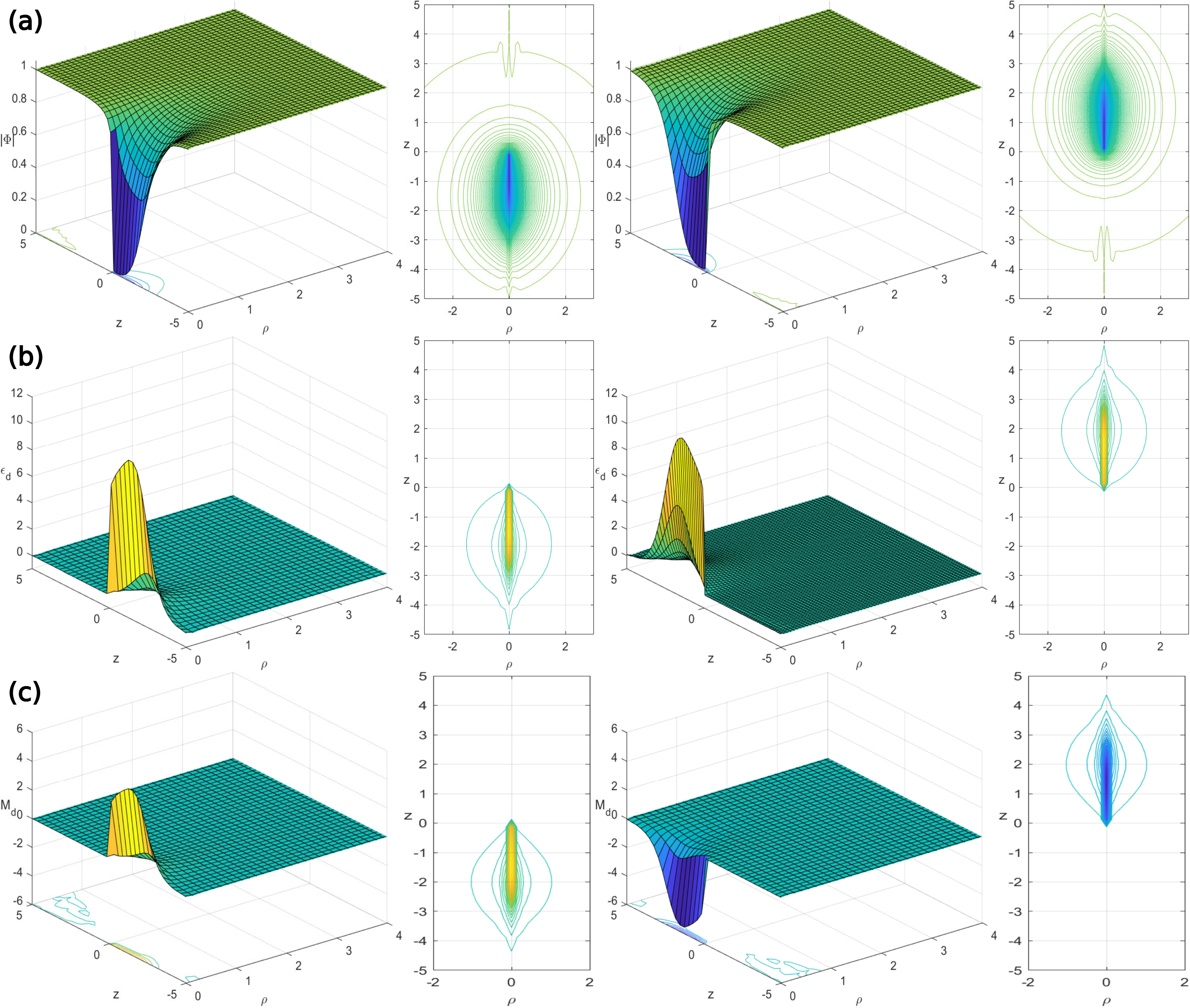}
\end{center}
\caption{(a) Higgs modulus; (b) energy density; and (c) magnetic charge density with their inset contour for Type I (left) and Type II (right) solutions for $n=1$, $\eta = 0$ at $\beta=0.7782$. }\label{compare}
\end{figure}

The Type II solution is constructed using the boundary conditions (\ref{boundary2}), (\ref{boundary1-negative}), (\ref{boundary3-negative}) and (\ref{boundary4-negative}). Our analysis reveals that the Type II half-dyon solutions possess magnetic charge $-2n\pi/g$ distributed along the positive $z$-axis, instead of negative $z$-axis as found in the Type I half-dyon. Despite the opposite magnetic charge, the Type II half-dyon solutions possess the same value for energy and electric charge as of Type I solutions. This shows that structure wise Type II half-dyon is a perfect reflection of Type I half-dyon about the $\rho=\sqrt{x^2+y^2}$ plane. The comparison between Type I and Type II half-dyon solutions for $n=1$, $\eta =0$ are shown in Fig.~\ref{compare}.

The Type II solutions have the same $\eta_{max}$ for $n=1,2,3$ and 4 as Type I solutions. Similarly, no solutions exist for Type II solutions for $n>4$. The existence of Type I and Type II solutions agree with the observation in Refs.~\citen{kn:18,kn:20,kn:32} where properties of monopole-antimonopole pair were studied. These results suggest that the magnetic monopole (or dyon) could appear in pair, indicating possibility of monopole pair production.

\section{Conclusions}\label{conclusions}

We have studied generalized half-dyon configurations in SU(2) YMH theory considering $\phi$-winding number $1 \leq n \leq 4$, Higgs self-coupling constant $0\leq \beta \leq 1$ and electric charge parameter $0\leq \eta <\eta_{max}$. For $ n = 1$, 2 and 3, $\eta_{max} = 1$ whereas for $n = 4$, $\eta_{max} = 0.9354$. These axially symmetric solutions are categorized as Type I and Type II half-dyon systems.

The Type I half-dyon is a finite length object carrying magnetic charge $+2n\pi/g$ and electric charge $Q(\eta)$ that extends from the origin and distributes along the negative $z$-axis. The Type II half-dyon on the other hand is a perfect reflection of Type I half-dyon over the origin, where it possesses magnetic charge $-2n\pi/g$ and distributed along the positive $z$-axis. Both Type I and Type II half-dyon possess same electric charge $Q(\eta)$ and dipole moment, despite having opposite value of magnetic charge.

Higher value of $n$ indicates superposition of multiple half-dyons. With higher $n$ and $\eta$, the shape of Type I half-dyon stretches further along the negative $z$-axis while the Type II half-dyon stretches further along the positive $z$-axis. However, the effect of stretching is much significant due to increase in $\phi$-winding number $n$ than increase in electric charge parameter $\eta$.

It is also interesting to examine the Type I half-dyon with $n=2$ and $\eta=0$. This two half-monopole configuration has total magnetic charge $4\pi/g$ and it is fundamentally a 't Hooft-Polyakov monopole with axial symmetry. For comparison, the energy of 't Hooft-Polyakov monopole is 2.5057 whereas the energy of two half-monopole is 5.8240, both in unit of $2\pi\nu/g$. Hence our finding suggests that 't Hooft-Polyakov monopole can undergo stretching and possesses higher energy than its spherically symmetry state.

The existence of Type I and Type II solutions validates the observation in Refs.~\citen{kn:18,kn:20,kn:32} and suggests that magnetic monopole (or dyon) could appear in pair, which indicates possible magnetic monopole pair production. Since all the solutions reported in the SU(2) YMH theory have their counterparts in Weinberg-Salam model, works on translating the Type I and Type II half-dyon solutions in this paper to Weinberg-Salam model are ongoing and will be reported in a future paper.



\begin{thebibliography}{99}  




\bibitem[1]{kn:1} P.A.M. Dirac, {\it Proc. Roy. Soc. Lond. A} {\bf 133}, 60 (1931).

\bibitem[2]{kn:2} G. 't Hooft, {\it Nucl. Phy.} {\bf B79}, 276 (1974).

\bibitem[3]{kn:3} A.M. Polyakov, {\it JETP Lett.} {\bf 20}, 194 (1974) .

\bibitem[4]{kn:4} E.B. Bogomol'nyi and M.S. Marinov, {\it Sov. J. Nucl. Phys.} {\bf 23}, 357 (1976).

\bibitem[5]{kn:5} M.K. Prasad and C.M. Sommerfield, {\it Phys. Rev. Lett.} {\bf 35}, 760 (1975).

\bibitem[6]{kn:6} E.B. Bogomol'nyi, {\it Sov. J. Nucl. Phys.} {\bf 24}, 449 (1976).

\bibitem[7]{kn:7} C. Rebbi and P. Rossi, {\it Phys. Rev. D} {\bf 22}, 2010 (1980).

\bibitem[8]{kn:8} R.S. Ward, {\it Commun. Math. Phys.} {\bf 79}, 317 (1981).

\bibitem[9]{kn:9} P. Forgacs, Z. Horvath and L. Palla, {\it Phys. Lett. B} {\bf 99}, 232 (1981).
 
\bibitem[10]{kn:10} P. Forgacs, Z. Horvath and L. Palla, {\it Nucl. Phys. B} {\bf 192}, 141 (1981).

\bibitem[11]{kn:11} M.K. Prasad, {\it Commun. Math. Phys.} {\bf 80}, 137 (1981). 

\bibitem[12]{kn:12} M.K. Prasad and P. Rossi, {\it Phys. Rev. D} {\bf 24}, 2182 (1981).

\bibitem[13]{kn:13} E.J. Weinberg and A.H. Guth, {\it Phys. Rev. D} {\bf 14}, 1660 (1976).

\bibitem[14]{kn:14} Rosy Teh and K.M. Wong, {\it J. Math. Phys.} {\bf 46}, 082301 (2005).

\bibitem[15]{kn:15} Rosy Teh and K.M. Wong, {\it Int. J. Mod. Phys. A} {\bf 20}, 4291 (2005).

\bibitem[16]{kn:16} P.M. Sutcliffe, {\it Int. J. Mod. Phys. A} {\bf 12}, 4663 (1997).

\bibitem[17]{kn:17} C.J. Houghton, N.S. Manton and P.M. Sutcliffe, {\it Nucl.Phys. B} {\bf 510}, 507 (1998).

\bibitem[18]{kn:18} B. Kleihaus and J. Kunz, {\it Phys. Rev. D} {\bf 61}, 025003 (2000).

\bibitem[19]{kn:19} B. Kleihaus, J. Kunz, and Y. Shnir, {\it Phys. Lett. B} {\bf 570}, 237 (2003).
 
\bibitem[20]{kn:20} B. Kleihaus, J. Kunz, and Y. Shnir, {\it Phys. Rev. D} {\bf 68}, 101701 (2003).

\bibitem[21]{kn:21} B. Kleihaus, J. Kunz, and Y. Shnir, {\it Phys. Rev. D} {\bf 70}, 065010 (2004).

\bibitem[22]{kn:22} E. Harikumar, I. Mitra, and H.S. Sharatchandra, {\it Phys. Lett. B} {\bf 557}, 303 (2003).

\bibitem[23]{kn:23} R. Teh, B.L. Ng and K.M. Wong, {\it Mod. Phys. Lett. A} {\bf 27}, 1250233 (2012).

\bibitem[24]{kn:24} E. Witten, {\it Phys. Lett. B} {\bf 86},  283 (1979).

\bibitem[25]{kn:25} B. Julia and A. Zee, {\it Phys. Rev. D} {\bf 11}, 2227 (1975).

\bibitem[26]{kn:26} F.A. Bais and J.R. Primack, {\it Phys. Rev. D} {\bf 13}, 819 (1976). 

\bibitem[27]{kn:27} R. Teh, B.L. Ng and K.M. Wong, {\it J. Phys. G} {\bf 40}, 035004 (2013).

\bibitem[28]{kn:29} S. Coleman, Classical Lumps and Their Quantum Descendants, in Zichichi, A. (eds) {\it New Phenomena in Subnuclear Physics}, The Subnuclear Series, Vol.~13 (Springer, Boston, MA, 1977), p.~297.

\bibitem[29]{kn:30} L.D. Faddeev, Nonlocal, nonlinear and nonrenormalisable field theories, in {\it Proc. Int. Symp., Alushta} (Dubna: Joint Institute for Nuclear Research, 1976), p.~207.

\bibitem[30]{kn:31} L.D. Faddeev, Lett. Math. Phys. {\bf 1}, 289 (1976).

\bibitem[31]{kn:32} K.G. Lim, Rosy Teh and K.M. Wong, {\it J. Phys. G: Nucl. Part. Phys.} {\bf 39}, 025002 (2012).

\bibitem[32]{kn:33} J.J. Van der Bij and E. Radu, {\it Int. J. Mod. Phys. A} {\bf 17}, 1477 (2002).

\end{thebibliography}
\end{document}